\documentclass[reprint,superscriptaddress,amsmath,amssymb,showpacs,aps,pra]{revtex4-1}

\usepackage{graphicx}

\newcommand{\bra}[1]    {\langle #1\vert}
\newcommand{\ket}[1]    {\vert #1 \rangle}
\newcommand{\e}         {\mathrm{e}}
\newcommand{\ii}         {\mathrm{i}}

\begin{document}

\title{Noise assisted Ramsey interferometry}

\author{U.~Dorner}
\affiliation{Centre for Quantum Technologies, National University of Singapore, 117543, Singapore}
\affiliation{Clarendon Laboratory, University of Oxford, Parks Road, Oxford OX1 3PU, United Kingdom}

\date{\today}


\begin{abstract} 
  I analyze a metrological strategy for improving the precision of
  frequency estimation via Ramsey interferometry with strings of atoms
  in the presence of correlated dephasing. This strategy does not
  employ entangled states, but rather a product state which evolves
  into a stationary state under the influence of correlated
  dephasing. It is shown that by using this state an improvement in
  precision compared to standard Ramsey interferometry can be
  gained. This improvement is not an improvement in scaling, i.e. the
  estimation precision has the same scaling with the number of atoms
  as the standard quantum limit, but an improvement proportional to
  the free evolution time in the Ramsey interferometer.  Since a
  stationary state is used, this evolution time can be substantially
  larger than in standard Ramsey interferometry which is limited by
  the coherence time of the atoms.
\end{abstract}

\maketitle

% 03.67.Ac Quantum algorithms, protocols, and simulations 06.20.Dk
% Measurement and error theory 06.30.Ft Time and frequency 03.67.Pp
% Quantum error correction and other methods for protection against
% decoherence 42.50.St Nonclassical interferometry, subwavelength
% lithography

\section{Introduction}

Quantum enhanced precision measurements can drastically increase the
precision of sensing devices. The wide range of applications include,
for example, gravitational wave detectors, laser gyroscopes,
ultra-sensitive magnetic field detectors, and frequency estimation via
Ramsey interferometry which can potentially improve the precision of
atomic
clocks~\cite{glm04,Dowling08,Bollinger96,Chwalla07,Leibfried04,Roos06,Meyer01,Leroux10,Polzik10,Gross10,Riedel10}. Quantum
enhancement in such applications is generally achieved by preparing
the system in a quantum state which has a higher susceptibility with
respect to the quantity to be probed. In the absence of decoherence
this ideally improves the precision of the measurement device from the
standard quantum limit to the Heisenberg limit~\cite{Boixo07}. In
practical realizations, however, the presence of unwanted noise, which
threatens to destroy the coherence of the quantum states employed, has
to be taken into account. It is therefore of great importance to find
methods which are noise tolerant and simultaneously provide an
improvement in measurement precision~\cite{Dorner09,NTP11}.

In this paper I present such a method which can be used to improve the
precision of frequency estimation.  The method is based on Ramsey
interferometry in which a system consisting of $N$ two-level atoms
evolves freely in between two Hadamard gates where it picks up a phase
relative to a local oscillator. Depending on the transition frequency
$\omega$ of the atoms the local oscillator is typically either a laser
or a microwave field. The measurement of the internal state of the
atoms is then used to estimate $\omega$ with a statistical uncertainty
$\Delta\omega$ which we want to be as small as possible.
Unfortunately, the presence of unavoidable experimental noise
typically increases the estimation uncertainty $\Delta\omega$. Here, I
discuss a situation where correlated dephasing is a significant source
of noise which is the case in recent experiments with strings of
trapped
ions~\cite{Roos06,Chwalla07,Monz10,Langer05,Kielpinski01,Leibfried04}. Under
these circumstances, an initial product state of the atoms evolves
into a stationary, mixed state which has been experimentally prepared
and studied using two atoms~\cite{Chwalla07}. But here I go beyond
$N=2$ and present a method to employ these states for improving the
precision of frequency estimation. In particular, I show that, in the
absence of any further experimental imperfections, the precision
$\Delta\omega$ behaves like in a completely noise-less system, thus
significantly improving $\Delta\omega$ compared to standard Ramsey
interferometry. I then extend this approach by taking into account
further, relevant experimental imperfections. Specifically, I discuss
the effect of (i) imperfect gate operations, (ii) imperfect
measurements, and (iii) spontaneous emission. Taking these into
account I show that the method discussed in this paper can lead to
improvements in measurement precision of one order of magnitude
compared to standard Ramsey interferometry.

Ideas to use quantum states to reduce $\Delta\omega$ in Ramsey
interferometry go back to Bollinger et al.~\cite{Bollinger96} in which
a system consisting of $N$ two-level atoms, e.g. a string of ions
stored in a Paul trap, is prepared in a multi-particle Greenberger
Horne Zeilinger (GHZ) state $\ket{0\ldots0}+\ket{1\ldots1}$
($\ket{0},\,\ket{1}$ denoting the two internal states). Using this
state in a Ramsey interferometer instead of the product state
$(\ket{0}+\ket{1})^{\otimes N}$ reduces $\Delta\omega$ by a factor
$\sqrt{N}$. That is, the estimation uncertainty is reduced from the
standard quantum limit ($\Delta\omega\sim 1/\sqrt{N}$) to the
Heisenberg limit ($\Delta\omega\sim 1/N$). However, it has been shown
subsequently by Huelga et al.~\cite{Huelga97} that this gain in
precision is completely annihilated if a particular type of noise is
taken into account. In ~\cite{Huelga97} this noise was assumed to be
uncorrelated, Markovian dephasing, i.e. each atom dephases completely
independently from all other atoms, and the fluctuations causing the
dephasing are Markovian.  Interestingly, if the Markov assumption is
dropped, but the dephasing of different atoms is still uncorrelated,
recent works of Matsuzaki et al. and Chin et al. have shown that it is
still possible to beat the standard quantum limit when using entangled
states~\cite{Matsuzaki11,Chin12}. In contrast to this, in this paper I
focus on correlated dephasing caused by fluctuating magnetic fields
(and not fluctuations of the local oscillator) as it occurred in
recent trapped ion
experiments~\cite{Roos06,Chwalla07,Monz10,Langer05,Kielpinski01,Leibfried04}. The
dephasing is correlated since the magnetic field (and its
fluctuations) are the same for all ions.  It was shown previously that
in this case non-classical states which are elements of decoherence
free subspaces have a significantly improved coherence
time~\cite{Monz10,Roos06,Langer05,Kielpinski01,Roos05,Dorner12}. In~\cite{Dorner12},
which extends on a method first presented in~\cite{Roos05,Roos06}, it
was shown that by a suitable choice of internal states of the ions it
can be arranged that a GHZ state is decoherence free and
simultaneously improve the precision of frequency estimation by a
factor $\sqrt{N}$. The major technical difficulty in that method is to
prepare a GHZ state for large $N$. Dramatic experimental improvements
have been made in this respect during recent
years~\cite{Leibfried05,Monz10}, the current record being a fidelity
of $50.8\%$ for $N=14$ ions~\cite{Monz10}. Despite these achievements,
it is still very challenging to prepare a large GHZ state. In this
paper I therefore present a method for reducing $\Delta\omega$ which
does not require the preparation of a GHZ state (or any other
entangled state) but only a product state which is experimentally much
less demanding. The key idea of the method is to employ two different
transitions within a string of atoms which dephase in an
anti-correlated manner under the influence of correlated
fluctuations~\cite{Roos05} (see Fig.~\ref{fig1}). An initial product
state $(\ket{0}+\ket{1})^{\otimes N}$ then evolves into a stationary
state which is used for Ramsey interferometry.  This improves the
estimation uncertainty not in terms of scaling compared to
conventional Ramsey interferometry, i.e. still $\Delta\omega\sim
1/\sqrt{N}$, but by a factor proportional to $\sqrt{\gamma
  t}$, where $t$ is the free evolution time in the Ramsey
interferometer and $1/\gamma$ is the single-atom coherence time. Since
the method employs a stationary state, $t$ can be substantially larger
than in Ramsey interferometry with single atoms and will mainly be
limited by spontaneous decay of the atoms.  Taking this, as well as
imperfect gates and measurements into account, I will show that
improvements of one order of magnitude in measurement precision are
feasible.

\section{System and noise model}
\label{sec:setup}

\begin{figure}[t]
  \centering\includegraphics[]{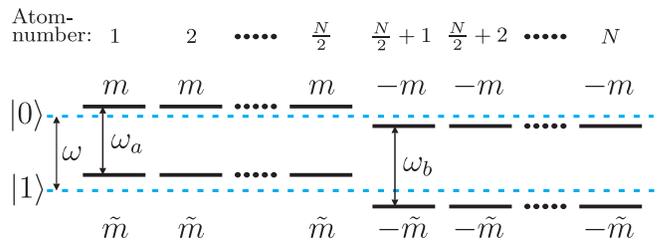}
  \caption{%
    The system under consideration consists of $N$ two-level atoms,
    where the two states of half of the atoms have magnetic quantum
    numbers $m$ and $\tilde m$ and the other half have magnetic
    quantum numbers $-m$ and $-\tilde m$. A magnetic field leads to
    linear Zeeman shifts which are equal in magnitude but opposite in
    sign. The goal is to precisely estimate the transition frequency
    $\omega$.}
  \label{fig1}
\end{figure}

The system under consideration consists of $N$ two-level atoms with
internal states $\ket{0}$ and $\ket{1}$ as shown in
Fig.~\ref{fig1}. The two internal states of half of the atoms are
required to have magnetic quantum numbers $m$ and $\tilde m$ and the
other half $-m$ and $-\tilde m$. In addition, all upper states are
elements of the same Zeeman manifold and all lower states are elements
of the same Zeeman manifold, and the two manifolds are separated by a
frequency $\omega$. Applying a sufficiently weak magnetic field leads
to linear Zeeman shifts of the atomic levels which are, due to the
choice of magnetic quantum numbers, equal in magnitude but opposite in
sign for the two cases, i.e. $\omega_a=\omega-\varepsilon$ and
$\omega_b=\omega+\varepsilon$. An immediate consequence of this is
that the transition frequency $\omega$, which we aim to measure, is
given by $\omega = (\omega_a+\omega_b)/2$, and, by construction, is
independent of the magnetic field. The frequency $\omega$ is therefore
similar to a 'clock transition'. Fluctuations of the magnetic field
leads, again due to the linear Zeeman effect, to fluctuations in the
frequencies $\omega_a$ and $\omega_b$ which are again equal in
magnitude but have opposite sign.  If the two-time correlation
function of these fluctuations decays faster than any other relevant
time scale in the system it can be approximated by a delta function
(see~\cite{Dorner12} for a detailed derivation) which leads to a
Markovian master equation describing the system dynamics,
\begin{equation}
  \dot\rho = -\ii[H,\rho] + \frac{\gamma}{2} 
  \left( L\rho L -\frac{1}{2}L^2\rho - \frac{1}{2}\rho L^2 \right).
  \label{eq:eq_of_motion}
\end{equation}
Here, $H$ is the system Hamiltonian, $\gamma$ is a dephasing rate and
\begin{equation}
  L=-\sum_{j\le N/2}\sigma_z^j + \sum_{j> N/2}\sigma_z^j,
  \label{eq:Hamil}
\end{equation}
where $\sigma_z^j$ is the Pauli $z$-operator acting on atom $j$. This
system is used to perform Ramsey interferometry, 
i.e. the system is initially prepared in a product state
\begin{equation}
  \ket{\psi_{in}} = \left( \frac{1}{\sqrt{2}}\left( \ket{0}+\ket{1} \right)\right)^{\otimes N}
  \label{eq:ini_state}
\end{equation}
by applying a Hadamard gate to all atoms using two light fields
(lasers or microwaves) of frequency $\omega_{L,a}$ and $\omega_{L,b}$
which might be slightly detuned from the atomic transition frequencies
$\omega_a$ and $\omega_b$. In practice, the Hadamard gate would be
realized by performing a $\pi/2$-pulse which is, strictly speaking,
not the same as a Hadamard gate but the difference has no effect on
the estimation uncertainty of $\omega$.  After the Hadamard gate, the
Hamiltonian of the system in a rotating frame takes the form
\begin{equation}
  H = \frac{\delta_a}{2}\sum_{j\le N/2}\sigma_z^j  + \frac{\delta_b}{2}\sum_{j> N/2}\sigma_z^j
\end{equation}
with $\delta_i\equiv \omega_i - \omega_{L,i}$, $i=a,b$.  The system
then evolves according to Eq.~(\ref{eq:eq_of_motion}) for a `free
evolution' time $t$ before a second Hadamard gate and a measurement of
the atoms in the $\{\ket{0},\,\ket{1}\}$-basis is performed. In this
paper it is assumed that $t$ is sufficiently large to let the system
evolve into a stationary state.  The whole procedure is repeated $\nu$
times leading to $\nu$ measurement results which provide, by using an
appropriate estimator, an estimate $\omega_{est}$ of the transition
frequency $\omega$. The statistical uncertainty $\Delta\omega$ of
$\omega_{est}$ can be quantified by~\cite{Braunstein94,Braunstein96}
\begin{equation}
  \Delta\omega = 
  \left\langle \left( \frac{\omega_{est}}{|\partial\langle \omega_{est}\rangle/\partial\omega|}-\omega  \right)^2  \right\rangle^{1/2}
  \label{eq:uncertainty}
\end{equation}
which, in the case of an unbiased estimator, is simply the standard
deviation. The uncertainty, or precision, $\Delta\omega$ is bounded
from below by the Cram\'er-Rao bound and the quantum Cram\'er-Rao
bound~\cite{Helstrom,Braunstein94,Braunstein96}
\begin{eqnarray}
  \Delta\omega &\ge& \frac{1}{\sqrt{\nu F}} = \frac{1}{\sqrt{T F/t}} \equiv \Delta\omega_{CR}  \\
  &\ge& \frac{1}{\sqrt{\nu F_Q}} = \frac{1}{\sqrt{T F_Q/t}}  \equiv \Delta\omega_{QCR},
  \label{eq:crb}
\end{eqnarray}
where $F$ is the Fisher information and $F_Q$ is the quantum Fisher
information (QFI).  Furthermore, $T\equiv\nu t$ is the total time of the
experiment, where $t$ is the time needed for a single
experimental run which for simplicity is assumed to be 
approximately equal to the free evolution time between the two
Hadamard gates.  The Fisher information $F$ depends on the state of
the system before the measurement and the particular measurement we
perform while the QFI depends only on the state before the
measurement.  The first bound in Eq.~(\ref{eq:crb}) can be reached via
maximum likelihood estimation for large $\nu$ and the second bound by
an {\em optimal} measurement which always exists~\cite{Braunstein94}.

A word of caution is in order if the above model
[i.e. Eq.~(\ref{eq:eq_of_motion})] is applied to the experimental
setup described in~\cite{Monz10} since there it has been pointed out
that the magnetic field fluctuations leading to dephasing are
non-Markovian.  Nonetheless, it should be possible to apply the method
described in this paper to the case~\cite{Monz10} since it only relies
on the stationary state (i.e. $\gamma t\gg1$).  This state will
consist of an incoherent mixture of eigenstates of $L$ which will be
the same for the Markovian and the non-Markovian case
of~\cite{Monz10}. Furthermore, we can always artificially enforce the
noise to be of the form~(\ref{eq:eq_of_motion}) such that the initial
state relaxes quickly into the stationary state. Any additional
non-Markovian fluctuations of the magnetic field should then have no
effect.

\section{Benchmarks}
\label{sec:bench}

In conventional Ramsey interferometry, i.e. if the {\em same} internal
states for all atoms are used, and in the absence of magnetic field
fluctuations, the best possible precision in case of a product
state~(\ref{eq:ini_state}) turns out to be $1/\sqrt{TtN}$, i.e. the
standard quantum limit. Using an $N$-particle GHZ state
$\ket{\psi_{in}^{GHZ}}= ( \ket{0}^{\otimes N} + \ket{1}^{\otimes N} )
/\sqrt2$ instead would ideally improve this precision to $1/\sqrt{Tt}
N$, i.e. the Heisenberg limit~\cite{Bollinger96}. However, if magnetic
field fluctuations are taken into account this improvement is
diminished significantly. In fact, if uncorrelated, Markovian
dephasing with dephasing rate $\gamma$ is assumed it has been shown
that the best possible precision in Ramsey interferometry is given by
\begin{equation}
  %\Delta\omega_{CR}^{(1)}=\Delta\omega_{QCR}^{(1)}
\Delta\omega_{bench} = \sqrt{ \frac{2\gamma \e}{NT} }
  \label{eq:prod_prec}
\end{equation}
for {\em both} the product state~(\ref{eq:ini_state}) and the GHZ
state, i.e. the two states are metrologically
equivalent~\cite{Huelga97}. It should be noted that
Eq.~(\ref{eq:prod_prec}) is based on the quantum Cram\'er-Rao bound
which in this case is equal to the Cram\'er-Rao bound. Furthermore, in
order to obtain expression~(\ref{eq:prod_prec}), an optimal free
evolution time $t=t_{opt}=1/2\gamma$ in case of a product state and
$t=t_{opt}=1/2N\gamma$ in case of a GHZ state has been assumed.

In the presence of correlated dephasing as in
Eq.~(\ref{eq:eq_of_motion}) the situation gets even worse. In fact, it
has been calculated numerically in~\cite{Dorner12} that for a product
state~(\ref{eq:ini_state}) the precision, i.e. the quantum
Cram\'er-Rao bound, is given by
\begin{equation}
  % \Delta\omega_{QCR}^{(2)}
  \Delta\tilde\omega_{bench} \approx (1.41 + 0.87/N^{0.90})\sqrt{\gamma/T}, 
  \label{eq:prod_prec2}
\end{equation}
and for a GHZ state a precision of $\sqrt{2\e\gamma/T}$ is obtained,
none of which tend to zero for large $N$~\cite{Dorner12}. A solution
to this problem was developed in~\cite{Dorner12}, where it was shown
that using a level scheme as in Fig.~\ref{fig1} and a GHZ state as
input of the Ramsey interferometer the precision is given by
$1/\sqrt{Tt} N\xi$, where $\xi$ is the preparation fidelity of the GHZ
state. Although incredible progress has been made to create large GHZ
states in ion traps~\cite{Monz10} it is still a very challenging and
expensive task to prepare such states. In the next sections I
therefore present a method which requires merely the product
state~(\ref{eq:ini_state}) which is easy to prepare in experiments. I
will use Eq.~(\ref{eq:prod_prec}) as a benchmark to measure the
performance of the method since the corresponding scenario employs
product states as well. Equation~(\ref{eq:prod_prec2}) will serve as a
benchmark to a lesser extend since this precision is considerably
worse than~(\ref{eq:prod_prec}). A situation
where~(\ref{eq:prod_prec2}) occurs as precision would therefore be
avoided in practice. The precision of the method discussed in the next
sections is not as good as $1/\sqrt{Tt} N\xi$, which relies on
entanglement, but clearly beats the benchmarks~(\ref{eq:prod_prec})
and~(\ref{eq:prod_prec2}).

\section{Estimation precision}
\label{sec:estimation_precision}

The method described in this paper relies on the stationary state
resulting from the dynamics described by Eq.~(\ref{eq:eq_of_motion})
given that the input state has the form~(\ref{eq:ini_state}). In the
following I will first derive an expression for this state and then
calculate the corresponding estimation uncertainties
$\Delta\omega_{CR}$ and $\Delta\omega_{QCR}$. To ease notation, I will
call the first $M=N/2$ atoms subsystem $A$ and the second $M$ atoms
subsystem $B$. Additionally, it is helpful to realize that
Eq.~(\ref{eq:eq_of_motion}) and the state~(\ref{eq:ini_state}) are
completely symmetric under particle exchange on subsystem $A$ and $B$,
respectively. Therefore, a Fock representation can be introduced, 
\begin{align}
  &  \ket{k,M-k}_A\ket{l,M-l}_B =  \nonumber\\
  &\frac{1}{\sqrt{c^M_k c^M_l}} \sum_P P \ket{i_1,i_2,\ldots,i_M}
  \sum_P P \ket{i_{M+1},i_{M+2},\ldots,i_N},
  \label{eq:fockstate}
\end{align}
where $c^M_k \equiv \binom{M}{k}$ and $i_j=0,1$ and $k$ ($N-k$) is the
number of zeros (ones) in $\ket{i_1,i_2,\ldots,i_M}$ and analogously
for $\ket{i_{M+1},i_{M+2},\ldots,i_N}$. Furthermore, both sums are
over all permutations $P$ of particles which lead to different terms
in each sum. Hence, the state $\ket{k,M-k}_A\ket{l,M-l}_B$ is a state
with $k$ atoms in state $\ket{0}$ and $M-k$ atoms in state $\ket{1}$
on subsystem $A$ and $l$ atoms in state $\ket{0}$ and $M-l$ atoms in
state $\ket{1}$ on subsystem $B$.  To make the notation simpler, in
the following I will use the abbreviations
$\ket{k}_A\equiv\ket{k,M-k}_A$ and $\ket{l}_B\equiv\ket{l,M-l}_B$.  In
this representation the product state~(\ref{eq:ini_state}) is given by
\begin{equation}
  \ket{\psi_{in}} = \frac{1}{2^M}\sum_{k,l=0}^M \sqrt{c^M_kc^M_l} \ket{k}_A\ket{l}_B .
\end{equation}
In addition, the operators $\sum_{j\le N/2}\sigma_z^j = a_0^\dagger a_0
- a_1^\dagger a_1$ and $\sum_{j> N/2}\sigma_z^j = b_0^\dagger b_0 -
b_1^\dagger b_1$ can be introduced, where $a_i^\dagger$ ($a_i$) are bosonic creation
(annihilation) operators of an atom in state $\ket{i},\,i=0,1$ in
subsystem $A$, and $b_i^\dagger$, $b_i$ act in the same way on
subsystem $B$. Using the fact that $a_0^\dagger a_0 + a_1^\dagger a_1
= b_0^\dagger b_0 + b_1^\dagger b_1=M$ the equation of motion
is of the form~(\ref{eq:eq_of_motion}) but now with
\begin{equation}
  L=2(b_0^\dagger b_0 - a_0^\dagger a_0)
\end{equation}
and
\begin{equation}
  H =\delta(a_0^\dagger a_0 + b_0^\dagger b_0) + \tilde\delta(b_0^\dagger b_0 - a_0^\dagger a_0),
\end{equation}
where
\begin{eqnarray}
  \tilde\delta &\equiv& \frac{\omega_b-\omega_a}{2} - \frac{\omega_{L,b}-\omega_{L,a}}{2},\\
  \delta &\equiv& \omega - \frac{\omega_{L,a}+\omega_{L,b}}{2}.
\end{eqnarray}
The state of the system at time $t$ is then given by
\begin{align}
  \rho(t) = &\e^{-\ii H t}
  \e^{-\frac{\gamma}{4}L^2t} \nonumber\\
  &\times\sum_{m=0}^\infty \frac{(\gamma t/2)^m}{m!} L^m
  \ket{\psi_{in}}\bra{\psi_{in}} L^m \e^{-\frac{\gamma}{4}L^2t}
  \e^{\ii H t} \nonumber\\
  =&\frac{1}{2^N}\sum_{k,l,j,n=0}^M  \sqrt{c^M_kc^M_lc^M_jc^M_n}  \ket{k}_A\ket{l}_B \bra{j}_A\bra{n}_B   \nonumber\\
  &\times \e^{-\gamma t (l-k+j-n)^2} \e^{-\ii\tilde\delta
    t(l-k+j-n)}\e^{-\ii\delta t(k+l-j-n)}
\end{align}
and the stationary state, i.e. $\gamma t\gg1$, is therefore
\begin{align}
  \rho_{stat} =&\frac{1}{2^N}\sum_{k,l,j,n=0}^M  \sqrt{c^M_kc^M_lc^M_jc^M_n}  \ket{k}_A\ket{l}_B \bra{j}_A\bra{n}_B   \nonumber\\
  &\times\e^{-\ii\delta t(k+l-j-n)}\delta_{l+j,k+n}.
  \label{eq:rho_stat}
\end{align}
Note that this state does not depend anymore on $\tilde\delta$. 

In the atomic basis $\rho_{stat}$ takes the form 
\begin{equation}
  \rho_{stat} = \sum_{k=0}^N p_k \e^{-\ii Ht}\ket{\psi_k}\bra{\psi_k}\e^{\ii Ht}, 
\end{equation}
where the $\ket{\psi_k}$ are eigenstates of $L$ which are symmetric in
$A$ and $B$, respectively and the $p_k$ are the dimensions of the
corresponding eigenspaces divided by $2^N$. The correlated noise
removes all coherences of the initial state except for those which are
unaffected by the noise.  For example for $N=2$ this leads to
\begin{align}
 &\ket{\psi_0} = \frac{1}{\sqrt2}(\ket{00}+\ket{11}),\, %  \nonumber \\
 \ket{\psi_1} = \ket{01},\, \ket{\psi_2} = \ket{10}
\end{align}
with $p_0=1/2,\,p_1=p_2=1/4$, and for $N=4$ we have
\begin{align}
  &\ket{\psi_0} = \frac{1}{\sqrt6} ( \ket{0000} + \ket{0101} + \ket{0110} \nonumber \\
  &\qquad\qquad + \ket{1001} + \ket{1010} + \ket{1111}  )\nonumber  \\
  &\ket{\psi_1} = \frac{1}{2}\left( \ket{0100} + \ket{1000} + \ket{1101} + \ket{1110}  \right) \nonumber \\
  &\ket{\psi_2} = \frac{1}{2}\left( \ket{1011} + \ket{0111} + \ket{0010} + \ket{0001}  \right) \nonumber \\
  &\ket{\psi_3} = \ket{1100},\,\ket{\psi_4} = \ket{0011}
\end{align}
and $p_0=3/8,\,p_1=p_2=1/4,\,p_3=p_4=1/16$. As can be seen, for $N=2$
the stationary state $\rho_{stat}$ is a Bell state with 50\%
fidelity. This state (except for a bit flip of the second atom), also
created by correlated dephasing, has been prepared in an experiment
where it has been demonstrated that it can lead to improvements in the
measurement of electric quadrupole moments and the line width of a
laser~\cite{Chwalla07}.

Based on $\rho_{stat}$ the QFI $F_Q$ can be calculated
(see Appendix~\ref{sec:AppII}) leading to
\begin{equation}
  \Delta\omega_{QCR}=\frac{1}{\sqrt{TtN}}
  \label{eq:QFI}
\end{equation}
which is the same expression as obtained for conventional Ramsey
interferometry with a product state in the {\em complete absence} of
dephasing. In other words, using the method discussed in this paper
would ideally eliminate all negative influences of magnetic field
fluctuations in the system.  This result is based on the QFI and
therefore does not necessarily correspond to the measurement scheme
performed in Ramsey interferometry, i.e. Hadamard gate and detection
of the state of the atoms. The performance of this particular
measurement can be studied by calculating the Fisher information which
is given by
\begin{equation}
  F = \sum_{k,l=0}^{\frac{N}{2}} \frac{1}{p(k,l|\omega)}\left(\frac{d}{d\omega}p(k,l|\omega)\right)^2,
  \label{eq:fisher}
\end{equation}
where $p(k,l|\omega)$ is the probability to find $k$ excited atoms in
$A$ and $l$ excited atoms in $B$ given that the value of the
transition frequency is $\omega$,
\begin{equation}
  p(k,l|\omega) = \bra{k}_A\bra{l}_B  H_g^{\otimes N} \rho_{stat}(\omega) H_g^{\otimes N} \ket{k}_A\ket{l}_B,
  \label{eq:probs}
\end{equation}
where $H_g$ is a Hadamard gate.
\begin{figure}[t]
  \centering\includegraphics[]{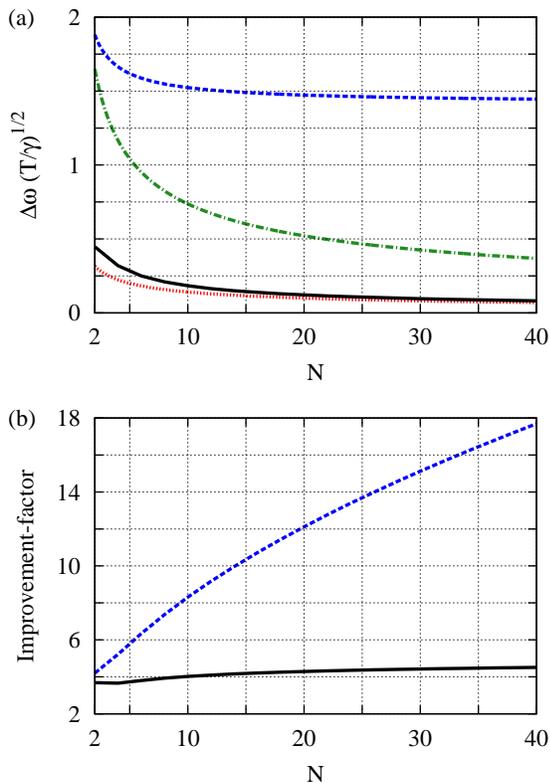}
  \caption{%
    (a) Estimation uncertainty, $\Delta\omega$, depending on the
    number of atoms $N$ for $\gamma t=5$. The solid (black) line is
    the estimation uncertainty $\Delta\omega_{CR}$ based on the Fisher
    information~(\ref{eq:fisher}). The dotted (red) line is the uncertainty
    $\Delta\omega_{QCR}$ based on the QFI and given by
    Eq.~(\ref{eq:QFI}). The dash-dotted (green) line and the dashed
    (blue) line are the benchmarks~(\ref{eq:prod_prec})
    and~(\ref{eq:prod_prec2}), respectively.  (b) Improvement factors
    $I$ (solid, black line) and $\tilde I$ (dashed, blue line)
    corresponding to the improvements over the
    benchmarks~(\ref{eq:prod_prec}) and~(\ref{eq:prod_prec2}),
    respectively.  }
  \label{fig2}
\end{figure}
From Eqs.~(\ref{eq:rho_stat}), (\ref{eq:fisher}) and~(\ref{eq:probs})
it is clear that the Fisher information has the form $F=t^2f(\delta
t)$. The quantity $f(\delta t)$ can then be calculated numerically and
maximized over $\delta t$, i.e.
\begin{equation}
  f_{max}\equiv \underset{\delta t}{\max} f(\delta t),
  \label{eq:fmax}
\end{equation} 
which yields $\Delta\omega_{CR}$. A result is given by the solid
(black) line in Fig.~\ref{fig2}(a). As can be seen, the uncertainty
$\Delta\omega_{CR}$ is slightly higher than the uncertainty based on
the QFI, Eq.~(\ref{eq:QFI}). This means that the measurement performed
in Ramsey interferometry is not optimal. However, the measurement
necessary to reach the precision~(\ref{eq:QFI}) will be a non-trivial
and, in general, non-local measurement which will be difficult to
implement (and therefore, in practice, will have finite fidelity). In
fact the maximum Fisher information follows approximately the
behavior $f_{max} = a_0 N + a_1$ where $a_0\approx
0.80\pm0.005,\,a_1\approx -2.24\pm0.12$ which is obtained by a fit to
data between $N=8$ and $N=40$ obtained by numerically calculating 
$f_{max}$. For large $N$ this approximately yields
\begin{equation}
  \Delta\omega_{CR} \approx \frac{1}{\sqrt{a_0TtN}}
  \label{eq:CR_stat_state}
\end{equation}
which is only slightly higher (approximately by a factor 1.1) than
Eq.~(\ref{eq:QFI}). It is therefore questionable if a more complex 
measurement is worth the effort.

It should be emphasized again that the results~(\ref{eq:fisher})
and~(\ref{eq:CR_stat_state}) are equal or similar to those obtained
for a completely noiseless system. Such an effect can, of course, also
be achieved by using states which are inherently insensitive to
magnetic fields, particularly `clock transitions' which use states
with zero magnetic quantum number. Therefore, the method presented in
this paper removes the restriction to clock transitions and makes it
possible to consider a greater variety of transitions for frequency
standard experiments.

A comparison of the precision $\Delta\omega_{CR}$ based on the Fisher
information~(\ref{eq:fisher}) and the benchmarks~(\ref{eq:prod_prec})
and~(\ref{eq:prod_prec2}) is shown in Fig.~\ref{fig2}(b). More
precisely, the solid (black) line shows the improvement factor
$I\equiv \Delta\omega_{bench}/\Delta\omega_{CR}$ and the dashed (blue)
line shows $\tilde I\equiv
\Delta\tilde\omega_{bench}/\Delta\omega_{CR}$ for $\gamma t=5$ which
means that the atoms can be kept 5 times longer than the coherence
time of a single atom. Unsurprisingly, $\tilde I$ grows with $N$ since
$\Delta\tilde\omega_{bench}$ approaches a constant, non-zero value for
large $N$. The improvement factor $I$ on the other hand, converges to
a constant value since both precisions scale like $\sqrt{N}$ for large
$N$. In particular, for $N\gg1$ the improvement factors are
\begin{eqnarray}
  I &\approx& \sqrt{2\e\gamma t}\sqrt{a_0+\frac{a_1}{N}}  \approx 2.09\sqrt{\gamma t}, \label{eq:I1}\\
  \tilde I &\approx& \left( 1.41 +\frac{0.87}{N^{0.90}}\right) \sqrt{ \gamma t (Na_0 +a_1)} 
  \approx 1.26\sqrt{N \gamma t}.\qquad \label{eq:I2}
\end{eqnarray}
Both improvement factors are proportional to $\sqrt{\gamma t}$ which
can be considerably larger than in conventional Ramsey interferometry
since there is no restriction due to the decoherence caused by magnetic field fluctuations.  In fact, it is the noise
which generates the state which is used and the improvement factor can
therefore be significant.

\section{Imperfect gates and measurements}
\label{sec:imperf}

In the previous sections imperfections of the 
Hadamard gates and imperfect measurements of the internal states of
the atoms have been neglected. An imperfect Hadamard gate can be
modeled by
\begin{equation}
  \mathcal{E}_H(\rho) = \eta_H H_g\rho H_g + \frac{1}{2}(1-\eta_H)\openone,
\end{equation}
where $H_g$ is a perfect Hadamard gate and $\eta_H$ characterizes the
probability to have a perfect gate. A faulty measurement of an atom
can be modeled using the measurement operators
\begin{equation}
  \Pi_i =\frac{1}{2}(1+\eta_M)\ket{i}\bra{i} +
  \frac{1}{2}(1-\eta_M)\sigma_x\ket{i}\bra{i}\sigma_x,
\end{equation}
where $i=0,1$, and $\eta_M$ is the likeliness that the correct
\begin{figure}[t]
  \centering\includegraphics[]{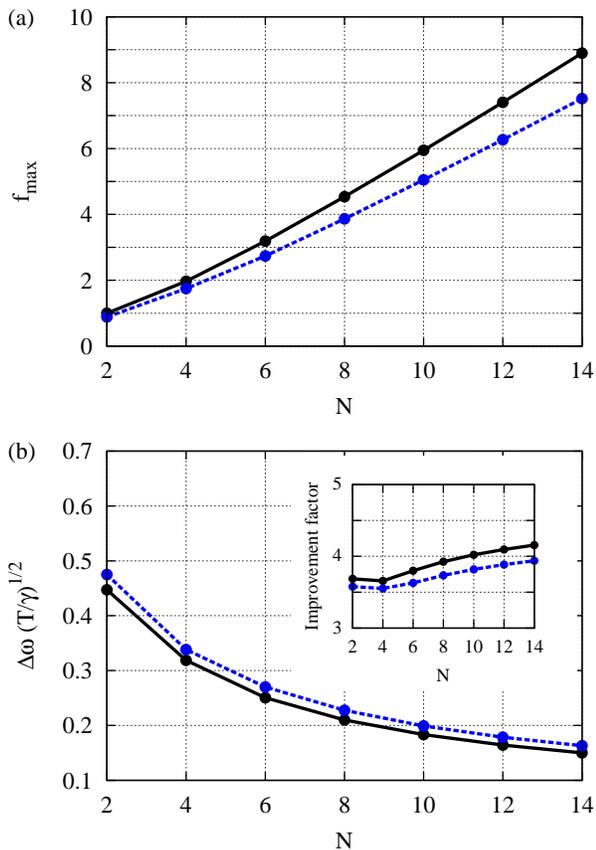}
  \caption{%
    (a) Maximum Fisher information $f_{max}$ defined by
    Eq.~(\ref{eq:fmax}) for $\eta_M=\eta_H=1$ (solid, black line) and
    $\eta_M=\eta_H=0.99$ (dashed, blue line) depending on the number
    of atoms $N$. (b) Estimation uncertainty $\Delta\omega_{CR}$ for
    $\eta_M=\eta_H=1,\,\gamma t=5$ (solid, black line) and estimation
    uncertainty $\Delta\omega_{CR}$ for $\eta_M=\eta_H=0.99,\,\gamma
    t=5$. Inset: Improvement factors $I$ (solid, black line) and
    $I_{\eta_M,\eta_H}$ (dashed, blue line).}
  \label{fig3}
\end{figure}
measurement result is obtained. Both measurement and Hadamard gate can
now be performed routinely with very high fidelities. Indeed, with
current ion trap technology gate and readout fidelities in excess of
$\eta_H=\eta_M\approx 0.99$ have been achieved~\cite{Lucas10}. Note
that errors in the state initialization preceding the first Hadamard
gate can be absorbed into these quantities. However, this
initialization, which is typically done via optical pumping, can be
done with very high fidelities even exceeding those of gate and
measurement. Furthermore, it should be emphasized that throughout this
paper it is assumed that exactly half of the atoms are in group $A$
and half of the atoms are in group $B$. This requires that group $A$
and $B$ can be addressed separately during the initialization phase
which can be easily achieved with current ion trap technology.
Despite the fact that $\eta_H$ and $\eta_M$ are close to one it has
been shown previously that they can have a significant effect on the
overall estimation precision. For example, in the method discussed
in~\cite{Dorner12} which relies on highly entangled GHZ states, these
imperfections increase the estimation uncertainty by a factor
$(\eta_M\eta_H)^N$, i.e. exponentially with the number of
atoms. Fortunately, if separable states are used, it is to be expected
that these imperfections have a much smaller effect. For example, the
benchmark~(\ref{eq:prod_prec}), which is based on a product state, is
now given by $\Delta\omega_{bench}^{\eta_M,\eta_H} =
\sqrt{2\gamma\e/NT}/ \eta_M\eta_H^2$~\cite{Dorner12}. Hence it is
merely reduced by a factor $\eta_M\eta_H^2$ which is independent of
$N$ and therefore has only a minor effect if $\eta_H$ and $\eta_M$ are
close to $1$.

In order to study the effect of imperfect measurements and Hadamard
gates on the method discussed in this paper, numerical calculations of
$f_{max}$ in the atomic basis have been performed for $\eta_{H,M}$
different from $1$. An example is shown in Fig.~\ref{fig3} (dashed,
blue line) depending on $N$ and for $\eta_H=\eta_M=0.99$. The dashed
line shows an approximately linear behaviour with respect to $N$ for
$N\gtrsim8$, i.e. $f_{max}\approx \tilde a_0 N + \tilde a_1$, where
$\tilde a_0,\,\tilde a_1$ are approximately independent of $N$.  As a
reference, $f_{max}$ for $\eta_H=\eta_M=1$ is plotted as well (black
line) which corresponds to the black line in Fig.~\ref{fig2}(a) which
also shows a linear behaviour as discussed in
Sec.~\ref{sec:estimation_precision}.  This shows that the precision
still scales like the standard quantum limit, $\Delta\omega_{CR} \sim
1/\sqrt{\tilde a_0 N + \tilde a_1}$ and therefore $\eta_{M,H}$ have
only a minor effect on the overall estimation precision if they are
close to $1$. $\Delta\omega_{CR}$ is shown in Fig.~\ref{fig3}(b)
(dashed, blue line). The loss in precision for $\eta_H=\eta_M=0.99$ is
small: It is merely a factor $1.09$ for $N=14$. The improvement factor
$I_{\eta_M,\eta_H} \equiv
\Delta\omega_{bench}^{\eta_M,\eta_H}/\Delta\omega_{CR}$ is shown in
the inset of Fig.~\ref{fig3}(b) (dashed, blue line) together with $I$
(black line) which is the same as the black line in
Fig.~\ref{fig2}. For example, for $N=14$ the improvement factor is
reduced from $4.16$ to $3.94$.

\section{The effect of  spontaneous emission}

It was shown in the previous sections that the improvement factors
$I$ and $\tilde I$ increase with $\sqrt{\gamma t}$. The value of $\gamma
t$ will be limited by a decay of the excited atomic state caused by
spontaneous emission with a decay time
$1/\Gamma$. References~\cite{Chwalla07} and~\cite{Monz10} report
$1/\gamma$ to be a few milliseconds and $8$ms, respectively (the
latter can be increased to $95\,$ms~\cite{Monz10}). The lifetime of
the excited atomic state in these experiments is $1/\Gamma =
1.17s$. For example, for $1/\gamma=5\,$ms the spontaneous decay time
is therefore $\gamma/\Gamma \approx 230$ times larger than the dephasing
time $1/\gamma$.
\begin{figure}[t]
  \centering\includegraphics[]{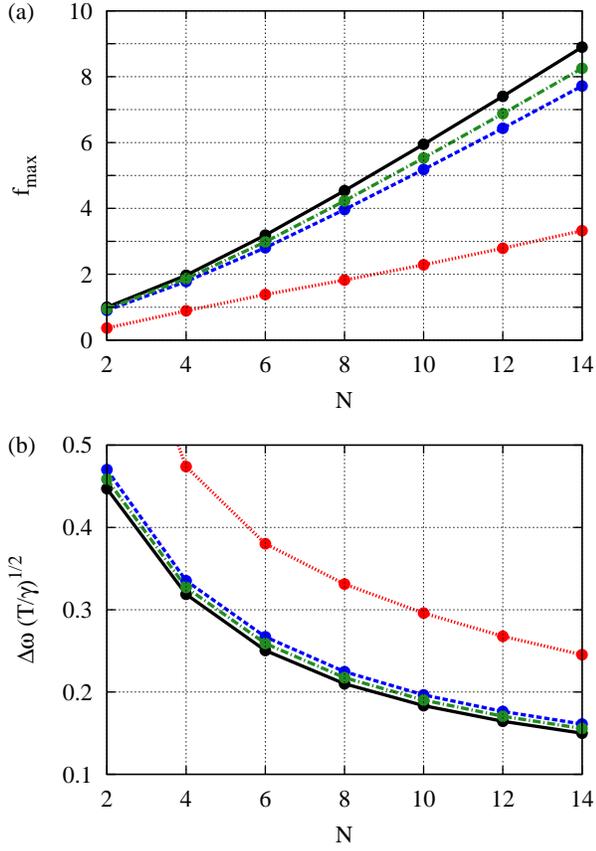}
  \caption{%
    (a) Maximum Fisher information $f_{max}$ [see Eq.~(\ref{eq:fmax})]
    for $\Gamma = 0$ (solid, black line), $\gamma/\Gamma =100$
    (dashed, blue line), $\gamma/\Gamma =200$ (dashed-dotted, green
    line) and $\gamma/\Gamma =10$ (dotted, red line) depending on the
    number of atoms $N$ ($\eta_M=\eta_H=1$ and $\gamma t =5$ in all
    cases). (b) Estimation uncertainty $\Delta\omega_{CR}$
    corresponding to the curves shown in (a).  }
  \label{fig4}
\end{figure}

Spontaneous emission can be included in our model by adding the term
\begin{equation}
  \Gamma \sum_{j=1}^{N}\left(\sigma_-^j \rho \sigma_+^j -\frac{1}{2}\sigma_+^j\sigma_-^j \rho -\frac{1}{2}\rho \sigma_+^j\sigma_-^j 
  \right)
\end{equation}
to the equation of motion~(\ref{eq:eq_of_motion}), where $\sigma_\pm =
(\sigma_x \pm \ii\sigma_y)/2$.  The fisher information $f_{max}$ can
then be calculated numerically and results are shown in
Fig.~\ref{fig4}(a) for $\gamma t = 5$ and $\gamma/\Gamma =200$
(dashed-dotted, green line), $\gamma/\Gamma =100$ (dashed, blue line)
and $\gamma/\Gamma =10$ (dotted, red line) ($\eta_M=\eta_H=1$). The
behavior is similar to that shown in Fig.~\ref{fig3}(a),
i.e. $f_{max}$ is approximately linear and the loss in precision is
approximately independent of $N$ for $N \gtrsim 8$. The corresponding
estimation uncertainties $\Delta\omega_{CR}$ are shown in
Fig.~\ref{fig4}(b). For example for $N=14$ the uncertainty is
increased by a factor $1.04$ for $\gamma/\Gamma =200$, $1.07$ for
$\gamma/\Gamma =100$ and $1.64$ for $\gamma/\Gamma =10$.

\begin{figure}[t]
  \centering\includegraphics[]{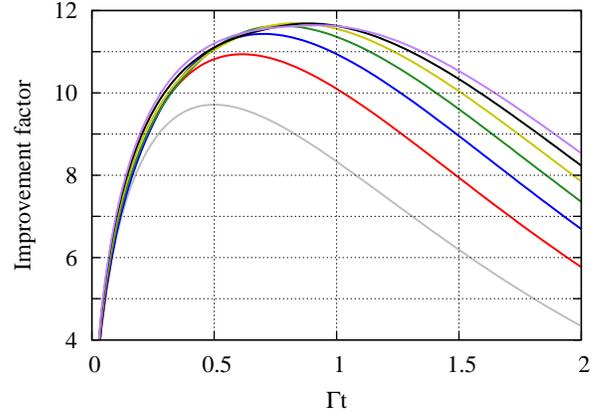}
  \caption{%
    Improvement factor $I_{\eta_M,\eta_H,\Gamma}$ given by
    Eq.~(\ref{eq:improv_spont}) versus $\Gamma t$ for $N=2,\ldots,14$
    (bottom to top) and $\eta_H=\eta_M=0.99,\,\gamma/\Gamma=200$.  }
  \label{fig5}
\end{figure}

Unsurprisingly, numerical simulations reveal that the Fisher
information $f_{max}$ decreases with increasing $\Gamma t$ but only
for $N=2$ does it exhibit an exponential decay ($f_{max}\sim
\exp(-2\Gamma t)$). For $N>2$ the decay is neither exponential nor
does it follow a power law. Taking into account spontaneous emission
and imperfect gates and measurements the
benchmark~(\ref{eq:prod_prec}) has to be modified to
$\Delta\omega_{bench}^{\eta_M,\eta_H,\Gamma} = \sqrt{(2\gamma+\Gamma)\e/NT}/
\eta_M\eta_H^2$. The improvement factor $I_{\eta_M,\eta_H,\Gamma}\equiv
\Delta\omega_{bench}^{\eta_M,\eta_H,\Gamma}/\Delta\omega_{CR}$ is then given by
\begin{equation}
  I_{\eta_M,\eta_H,\Gamma} = \sqrt{\frac{\gamma}{\Gamma}+\frac{1}{2}} \sqrt{\frac{2\e}{N}} \frac{1}{\eta_M\eta_H^2}\sqrt{\Gamma t f_{max}}.
\label{eq:improv_spont}
\end{equation}
Since $f_{max}$ is decreasing with $\Gamma t$ the term $\Gamma t
f_{max}$ has a maximum which corresponds to an optimal free evolution
time $t_{opt}$. Improvement factors for $N=2,\ldots,14$ (bottom to top) are
shown in Fig.~\ref{fig5}.
These examples take into account imperfect gates and measurements with
$\eta_H=\eta_M=0.99$. For $\gamma/\Gamma=200$, the maxima of the
curves yield optimal improvement factors $I_{\eta_M,\eta_H,\Gamma} =
9.7,\,10.93,\,11.42,\,11.62,\,11.68,\,11.67,\,11.63$ for
$N=2,\ldots,14$ corresponding to $\gamma t_{opt}=
100,\,123,\,140,\,155,\,167,\,177,\,186$. Note that the location of
the maxima, i.e. $\Gamma t_{opt}$, is independent of
$\gamma/\Gamma$. For $\gamma/\Gamma=10$ we get the optimal improvement
factors $I_{\eta_M,\eta_H,\Gamma} =
2.22,\,2.5,\,2.61,\,2.66,\,2.67,\,2.67,\,2.66$ for $N=2,\ldots,14$
corresponding to $\gamma t_{opt}=
5,\,6.2,\,7,\,7.8,\,8.4,\,8.9,\,9.3$. As can be seen, for example for
$\gamma/\Gamma=200$, the optimal improvement factor peaks at a value
of $11.68$ for $N=10$ and then decreases slightly. If this tendency
persists the best possible advantage over conventional Ramsey
interferometry is achieved for $N=10$. However, the behavior of
$I_{\eta_M,\eta_H,\Gamma}$ for $N>14$ is not known. In the worst case
the optimal improvement factor drops further. However, for the method
described in this paper to be worse than conventional Ramsey
interferometry, the optimal improvement factor would have to drop to a value
of below one. The precision is given by $\Delta\omega_{CR}= c
/\sqrt{N}I_{\eta_M,\eta_H,\Gamma}$ (with
$c=\sqrt{(2\gamma+\Gamma)\e/T}/\eta_M\eta_H^2$). Thus, if the optimal
$I_{\eta_M,\eta_H,\Gamma}$ does not change substantially for $N>14$,
the corresponding optimal precision decreases approximately with
$\sqrt{N}$.

\section{Conclusions}
I have shown that, by using a stationary state which is created by
correlated dephasing which represents a significant source of noise in
recent experiments with strings of trapped
ions~\cite{Monz10,Chwalla07,Roos06,Leibfried04,Langer05,Kielpinski01},
an improvement in the precision compared to standard Ramsey
interferometry can be gained.  This is due to the fact that the
measurement precision essentially behaves as in a completely noiseless
system, i.e. the precision is approximately given by
$1/\sqrt{TtN}$. On the other hand, in standard Ramsey interferometry
(under the influence of uncorrelated, Markovian dephasing) the free evolution
time is limited by the coherence time of the system, $t\sim 1/\gamma$,
leading to a precision proportional to $\sqrt{\gamma/NT}$. By
comparing the two situations it is obvious that an improvement
proportional to $\sqrt{\gamma t}$ can be gained.  This gain is
achieved at very low cost: The initial state is merely a product state
of $N$ atoms, created by optically pumping two groups of atoms into
two different internal states with magnetic quantum number $m$
and $-m$, and a subsequent Hadamard gate.

However, apart from correlated dephasing, further imperfections have
to be expected in practice. In particular, I took into account
imperfect Hadamard gates, imperfect measurements of the atomic states
and spontaneous emission. The effect of imperfect gates and
measurements is very small. It diminishes the estimation precision
merely by a constant factor independent of $N$ (approximately $1.1$
for $\eta_{H,M}=0.99$). The effect on the improvement factor, i.e. the
ratio of the precision of conventional Ramsey interferometry and the
precision of the method discussed in this paper, is small as well. For
example, for $\gamma t=5$ and $N=14$ atoms, the improvement factor is
reduced from 4.16 to 3.94. The effect of spontaneous emission is to
limit the free evolution time $t$ in the Ramsey interferometer.  The
examples shown in Figs.~\ref{fig2} and~\ref{fig3} assume $\gamma t=5$
which is a very conservative assumption. Recent experiments use
transitions with a spontaneous decay time exceeding
$1\,$s~\cite{Chwalla07} which is about 200 times larger than the
dephasing time. Under such circumstances I showed that, e.g. for
$N=10$ atoms, the free evolution time can be extend to $\gamma t=167$.
In conventional Ramsey interferometry the corresponding optimal
evolution time would be merely $\gamma t=0.5$. This leads to an
improvement in the estimation precision of one order of magnitude
compared to standard Ramsey interferometry.

A further source of noise are phase or frequency fluctuations of the
local oscillator. Since the local oscillator is the same for all
atoms, its fluctuations will effectively lead to correlated dephasing,
since all atoms are affected in the same way. This can be seen as
fluctuating frequency shifts of all atoms which are equal in magnitude
and have the same sign. This dephasing is of course independent of
magnetic quantum numbers and, for example, would also be present if
these are zero, i.e. if `clock-transitions' are used. However, laser
linewidths of about $\lesssim 1\,$Hz are now experimentally available
which renders the effect of this form of noise very
limited. Furthermore, a variation of the method presented in this
paper which is adapted from~\cite{Chwalla07,Roos06} could be used: If
all atoms fluctuate in the same way the stationary state would have
the same form as the one discussed in this paper except that half of
the atoms would be `spin-flipped'. As a consequence, however, the
quantity which can be estimated is not the arithmetic mean of two
frequencies but the difference between two frequencies.

The stationary state used in this paper is not entangled, however, it has recently
been pointed out that correlated noise can create quantum
discord~\cite{Lanyon13}. Whether the state considered in this paper
contains such non-classical correlations and whether they are the
reason for its superior performance in frequency estimation, is a
subject for future research.

\label{sec:conc}

\begin{acknowledgments}
  I acknowledge support for this work by the National Research
  Foundation and Ministry of Education, Singapore the Department of
  Atomic and Laser Physics, University of Oxford and Keble College,
  Oxford.
\end{acknowledgments}

\appendix

\section{Quantum Fisher information}
\label{sec:AppII}
Consider a system state $\rho$ which depends on a parameter $\omega$
which is to be estimated. The QFI is then given by~\cite{Helstrom,Holevo,Braunstein94,Braunstein96}
\begin{equation}
  F_Q = \sum_{j,k;\,p_k+p_j\ne0}\frac{2}{p_j+p_k}
\left |\left\langle\phi_j\left |\frac{d}{d\omega}\rho \right |\phi_k\right\rangle\right |^2.
\end{equation}
where $p_k$ and $\ket{\phi_k}$ are the eigenvalues and eigenvectors of $\rho$, respectively. The state $\rho_{stat}$ given by Eq.~(\ref{eq:rho_stat}) can be
diagonalized, 
\begin{equation}
  \rho_{stat} = \sum_{p=0}^M f_p \ket{\phi_p^0}\bra{\phi_p^0} + \sum_{p=0}^{M-1} f_p \ket{\tilde\phi_p^0}\bra{\tilde\phi_p^0} 
\end{equation}
with
\begin{eqnarray}
  f_p &=& \frac{1}{2^N}c^N_p, \\
  \ket{\phi_p^0} &=& \frac{1}{\sqrt{c^N_p}} \sum_{j=0}^p \sqrt{c^M_jc^M_{p-j}} \e^{-2\ii j\delta t}\ket{j}_A\ket{M-p+j}_B, \\
  \ket{\tilde\phi_p^0} &=& \frac{1}{\sqrt{c^N_p}} \sum_{j=0}^p \sqrt{c^M_jc^M_{p-j}} \e^{2\ii j\delta t}\ket{M-j}_A\ket{p-j}_B.\qquad\quad
\end{eqnarray}
The states $\ket{\phi_p^0},\,\ket{\tilde\phi_p^0}$ are orthogonal to
each other but to obtain a complete orthonormal basis set the 
states $\ket{\phi_p^k},\,\ket{\tilde\phi_p^k},\,k=1,\ldots,p$ have to be included 
which are eigenstates of $\rho_{stat}$ with eigenvalue $0$, i.e.
\begin{eqnarray}
  \langle\phi_p^k  | \phi_{p'}^{k'} \rangle &=& \delta_{pp'}\delta_{kk'} \\
  \langle\phi_p^k  | \tilde\phi_{p'}^{k'} \rangle &=& 0.
\end{eqnarray}
The quantum Fisher information then takes the form
\begin{align}
  F_Q = &4t^2\sum_{p=0}^{M} f_p  \sum_{k=1}^{p} \bra{\phi_0^p} a_0^\dagger a_0+b_0^\dagger b_0  \ket{\phi_k^p} \nonumber \\
  &+4t^2\sum_{p=0}^{M-1} f_p \sum_{k=1}^{p} \bra{\tilde\phi_0^p}
  a_0^\dagger a_0+b_0^\dagger b_0 \ket{\tilde\phi_k^p}.
\end{align}
The above expression can be evaluated leading to the simple result
$F_Q = Nt^2$ and therefore
\begin{equation}
  \Delta\omega_{QCR} = \frac{1}{\sqrt{\nu N} t} = \frac{1}{\sqrt{TtN}}.
\end{equation}

%\bibliography{article}
%Merlin.mbs v4.21 2009-07-09.
%

\end{document}